\documentclass[11pt]{article}
\usepackage{mathrsfs}
\usepackage{jheppub}
\usepackage{amsmath}
\usepackage{graphicx}
\usepackage{amsfonts}
\usepackage{amssymb}
\usepackage{flafter}
\usepackage{color}

\title{Dark Matter Induced Nucleon Decay: \\ Model and Signatures}

\author[a]{Junwu Huang}
\author[a,b]{and Yue Zhao}
\affiliation[a]{Stanford Institute of Theoretical Physics, Physics Department,\\
Stanford University, Stanford, CA 94305, USA}
\affiliation[b]{SLAC National Accelerator Laboratory, \\
2575 Sand Hill Road, Menlo Park, CA 94025, USA}

\emailAdd{curlyh@stanford.edu}
\emailAdd{zhaoyue@stanford.edu}

\abstract{ If dark matter (DM) carries anti-baryon number, a DM
particle may annihilate with a nucleon by flipping to anti-DM.
Inspired by Hylogenesis models, we introduce a single component DM
model where DM is asymmetric and carries B and L as -1/2.  It can
annihilate with a nucleon to an anti-lepton and an anti-DM at
leading order or with an additional meson at sub-leading order. Such
signals may be observed in proton decay experiments. If DM is
captured in the Sun, the DM induced nucleon decay can generate a
large flux of anti-neutrinos, which could be observed in neutrino
experiments. Furthermore, the anti-DM particle in the final state
obtains a relatively large momentum (few hundred MeV), and escapes
the Sun. These fast-moving anti-DM particles could also induce
interesting signals in various underground experiments.}

\begin{document}

\maketitle

\section{Introduction}\label{SEC::Intro}

Current astrophysical surveys and cosmological studies suggest that
dark matter (DM) constitutes about $27\%$ of the energy density of
the Universe~\cite{Ade:2013zuv}. The Standard Model (SM) of particle
physics cannot explain the abundance of the invisible component. New
fundamental physics are required to explain its existence and new
experiments are needed to study its nature.

Extensions beyond the standard models contain various weakly
interacting massive particles (WIMPs) that are candidates of
particle dark matter. In standard WIMP scenarios, the similar
magnitude of baryon and DM density, $ \Omega_{\mathrm{DM}}/ \Omega_B
\approx 5.5 $ is treated as a numerical coincidence. The baryon
asymmetry can be generated from CP-violating non-equilibrium
processes (such as the electroweak phase transition), while the DM
relic density is determined by thermal freeze out of WIMPs in the
early universe. The Asymmetric Dark Matter (ADM)
paradigm~\cite{Hut:1979xw,Nussinov:1985xr,Dodelson:1989cq,Gelmini:1986zz,Barr:1990ca,Barr:1991qn,Kaplan:1991ah,Kuzmin:1996he,Hooper:2004dc,Kaplan:2009ag},
however, provides a framework to relate the baryon and dark matter
density. For a review of Asymmetric Dark Matter models, please
see~\cite{Zurek:2013wia} and \cite{Petraki:2013wwa}. Dark Matter in
ADM models usually carries a conserved global charge shared by the
SM particles, namely, lepton or baryon number. Such a connection
naturally relates the number density of DM particles and SM
particles.  Therefore, the ADM paradigm naturally predicts dark
matter particles with a mass of a few GeV. The sensitivity of direct
detection experiments drop rapidly if the DM mass is small. New
unconventional signatures, if they exist, can help for the DM
search.

The existing ADM models generally fall into two classes depending on how the charge asymmetry is created:
\begin{itemize}
\item An initial charge asymmetry is first generated in either the visible or DM sector and later transferred to the other sector
by chemical equilibration through non-renormalizable operators. The
DM in this class of model carry the same baryon/lepton numbers as
the left-over SM particles.  Such scenario with different variations
is recently studied by
\cite{Hooper:2004dc,Kaplan:2009ag,Kribs:2009fy,Cai:2009ia,Cohen:2010kn,Shelton:2010ta,Haba:2010bm,Blennow:2010qp,McDonald:2010rn,Dutta:2010va,Falkowski:2011xh,Frandsen:2011kt}.
\item Equal and opposite charge asymmetries are generated via non-equilibrium CP-violating dynamics in the visible and DM sectors.
The DM in this class of models carries opposite charge baryon/lepton
numbers as the left-over SM particles. Such scenario is recently
studied by
\cite{Oaknin:2003uv,Kitano:2004sv,Kitano:2005ge,Farrar:2005zd,Gu:2007cw,Gu:2009yy,An:2009vq,An:2010kc,Gu:2010ft,Hall:2010jx,Heckman:2011sw,Kaplan:2011yj,Bell:2011tn,Cheung:2011if,MarchRussell:2011fi,Graesser:2011vj,Davoudiasl:2010am,Davoudiasl:2011fj,Blinov:2012hq}.
\end{itemize}

In this paper, we will focus on the second class of models. Very
recently, this class of models was re-visited
by~\cite{Davoudiasl:2010am,Davoudiasl:2011fj,Blinov:2012hq} and
interesting new experimental signatures were discussed. Let us first
review their model.  The interaction terms of the Lagrangian of this
model is written as follows (here we drop kinematic and mass terms
of the particles):

\begin{equation}
\mathcal{L} \supset \frac{\lambda_a}{M^2} \bar{X}_a \bar{d}_R^c
\bar{u}_R^c \bar{u}_R^c + \zeta_a \bar{X}_a \Psi^c \Phi^* + h.c.
\label{Eq:General}
\end{equation}
where $X_a, X_a^c (a = 1 , 2)$ are two vector pairs of hidden sector
fermions with masses $ m_{X_2}> m_{X_1} \geq \mathrm{TeV}$. $\Psi$
and $\Phi$ are two components of the DM relic in their model.  There
exists a physical CP-violating phase $\mathrm{arg}(\lambda_1^*
\lambda_2 \zeta_1 \zeta_2^*)$ that cannot be rotated away through
redefinition of the fields.

\begin{figure}[htb]
\centering
\includegraphics[width=0.6\columnwidth]{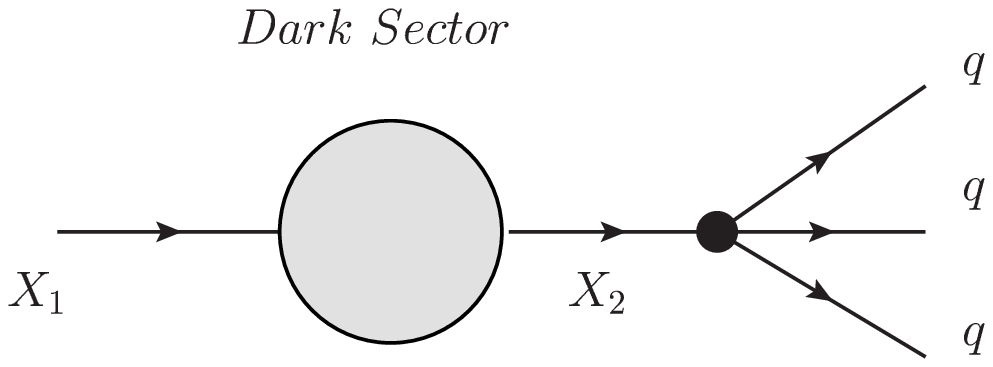}
\includegraphics[width=0.35\columnwidth]{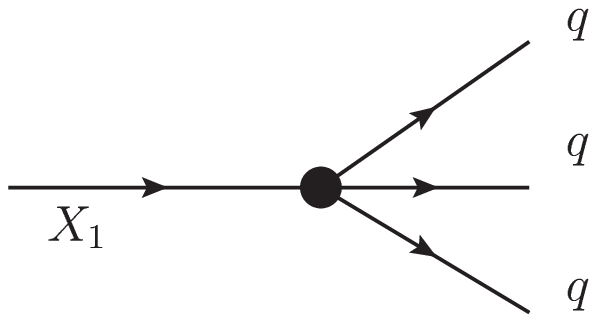}
\caption{Tree level and one-loop processes that generate an initial baryon asymmetry in the early universe}\label{Fig:Baryon}
\end{figure}

Baryogenesis begins when a non-thermal, CP- symmetric population of
$X_1$ and $\bar{X}_1$ is produced in the early Universe. As shown in
Fig.~\ref{Fig:Baryon} these states can decay to SM fields. The
interference between the two diagrams gives rise to an asymmetry
between the partial widths for $X_1 \rightarrow \bar{u} \bar{d}
\bar{d}$ and $\bar{X_1} \rightarrow udd$, while the same amount of
opposite asymmetry is deposited into DM sector.  The amount of
asymmetry induced through such interference is estimated as
\cite{Davoudiasl:2010am}
\begin{equation}
\epsilon = \frac{1}{2\Gamma_{X_1}}[\Gamma(X_1\rightarrow
udd)-\Gamma(\bar{X}_1\rightarrow\bar{u}\bar{d}\bar{d})]\simeq\frac{m^5_{X_1}
\textmd{Im}[\lambda^*_1\lambda_2\zeta_1
\zeta_2^*]}{256\pi^3|\zeta_1|^2 M^4 m_{X_2}}\label{Eq:asymmetry}
\end{equation}

To achieve successful baryogenesis, one needs to start with a proper
reheating temperature. It needs to be high enough to preserve
successful nucleosynthesis, while not too high to wash out the
asymmetry has already been generated.

As suggested
in~\cite{Davoudiasl:2010am,Davoudiasl:2011fj,Blinov:2012hq},
interesting signatures could be induced in such models. Since DM
particles carry anti-baryon number, they can annihilate visible
baryonic matter and produce meson in the final state, i.e. $\Phi +
N\rightarrow \bar{\Psi}+M$, where $N$ indicates nucleon and $M$ is
meson. Since dark matter particles are invisible, this signal mimics
the nucleon decay signatures and offers the possibility to search
for dark matter through nucleon decay experiments, e.g.
Super-Kamiokande (SuperK)~\cite{Nishino:2012ipa}. However, in the
model introduced in previous works, two species, one fermion and one
boson, of DM are necessary. If the two species of DM particles can
decay to each other, then the previously generated baryon asymmetry
would be washed out.  To avoid such decay, the splitting between
their masses needs to be smaller than the sum of proton and electron
mass.

In this paper, inspired by the work
of~\cite{Davoudiasl:2010am,Davoudiasl:2011fj,Blinov:2012hq}, we
present an alternative model which provides similar signatures as
induced nucleon decay (IND) processes. The model has the following
advantages and features which we will discuss in detail in the
following sections:

\begin{itemize}
\item We only have one species of dark matter particle, thus no degeneracy between DM masses is required.
\item Baryogenesis can be achieved in a similar manner as Hylogenesis.  If one assumes zero total baryon/lepton number to start with, we have a concrete prediction on DM mass.
\item We have an additional lepton in the final state.  This lepton helps to mimic the signatures of proton decay in the best constrained channel. The most sensitive channel in SuperK, $p^+ \rightarrow \pi^0 + e^+$ has the same visible final state particles as $p^+ + \phi \rightarrow \overline{\phi} + \pi^0 + e^+$, leading to a better experimental sensitivity of the IND signals in this model.
\item Besides to the 2-to-3 process studied in proton decay experiments, $p^+ + \phi \rightarrow \overline{\phi} + \pi^0 + e^+$, we also have
the leading order processes, $p^+ + \phi \rightarrow \overline{\phi} + e^+$
and $n + \phi \rightarrow \overline{\phi} + \bar{\nu}$. These processes have
much larger cross section, which can lead to significant anti-DM and
anti-neutrino fluxes from the Sun.  This can also be studied by
underground experiments such as SuperK.
\end{itemize}

The paper is organized as follows. In Sec.~\ref{SEC::Lag}, we
present our model Lagrangian and study its ultraviolet (UV)
completion. In Sec.~\ref{SEC::Parameters}, we discuss constraints on
model parameters from direct detection experiments and collider
physics and provide a benchmark set of parameters. In
Sec.~\ref{SEC::Chiral}, using the language of chiral Lagrangians, we
compute the cross sections of processes that provide interesting
experimental signatures. In Sec.~\ref{SEC::signature}, we provide a
systematic study of three interesting signatures in our model,
emphasizing both current constraints and future reaches.
Sec.~\ref{SEC::Conclusion} serves as a conclusion.

\section{Model Lagrangian and Parameter choices}\label{SEC::Lag}

In this section, we introduce a simple model with one component of
dark matter (DM).  The DM carries $-\frac{1}{2}$ unit of baryon
number and lepton number.  DM can annihilate with a proton/neutron
to an anti-lepton and an anti-DM in the final state. The baryon and
lepton numbers are still conserved in this process. Since the
DM/anti-DM in the initial/final state are not detected, such an
event fakes a nucleon decay event.

\subsection{Effective operators}\label{SEC::Effective}
Let us first write down the effective operators which lead to this
induced nucleon decay process.  First, since we only introduce one
species of DM particle in our model, a DM particle should be in the
initial state and an anti-DM is in the final state. Thus we need two
copies of DM field in the effective operator. Further, baryon number
is changed by one unit in the interaction, to preserve $SU(3)_c$ at
the meanwhile, at least three quarks are needed in the operator, and
the color indices should be anti-symmetrized.  Finally, to preserve
Lorentz symmetry, we need one more fermion field in the operator,
where the lepton field fits in. To make the operators we write down
have the lowest possible dimension, we choose our DM particle to be
a scalar field. Following the logic above, one can write down the
effective operators \footnote{There could be other choices, but here
we just list two typical ones.  The other choices of effective
operators will have the similar phenomenology.}
\begin{equation}
 \mathcal{O}_{S} = \frac{1}{\Lambda^4} \overline{\phi}^2(e^c u^c)(d^c u^c) \label{Eq:EffSinglet}
\end{equation}
or
\begin{equation}
 \mathcal{O}_{D} = \frac{1}{\Lambda^4} \overline{\phi}^2(L^\dag Q^\dag)(u^c d^c) \label{Eq:EffDoublet}
\end{equation}
The first operator $\mathcal{O}_{S}$ only involves $SU(2)_{W}$
singlet, so only charged lepton shows up in the final state.  The
second choice, $\mathcal{O}_{D}$, can generate either a charged
lepton or an anti-neutrino in the final state.  This will lead to
different signatures to search for.

The operators we write down are dimension 8.  One may worry whether
one can achieve a sizable signal rate with such high dimension
operators. However, this depends highly on the UV model which
generates these effective operators. We will address this issue in
the rest of this section and show that with a reasonable choice of
parameters, various experiments could probe interesting parameter
space of this model.

\subsection{UV completion of the Lagrangian}\label{SEC::Lagrangian}
Now let us go into more detail on how to realize these effective
operators.  As in
to~\cite{Davoudiasl:2010am,Davoudiasl:2011fj,Blinov:2012hq}, we
introduce the heavy particle $X$, and it couples to the quarks
through the following effective operators\footnote{Here we do not
try to UV complete this operator. Detailed discussion can be found
in~\cite{Davoudiasl:2010am,Davoudiasl:2011fj,Blinov:2012hq}.  In
later section, we will discuss the collider constraint on this
operator.}:

\begin{equation}
 \mathcal{O}_{Xq,S} =  \frac{1}{\Lambda^2} (X u^c)(d^c u^c) \label{Eq:XcolorS}
\end{equation}
for $\mathcal{O}_{S}$, and
\begin{equation}
 \mathcal{O}_{Xq,D} = \frac{1}{\Lambda^2} (X^c Q)(u^{c\dag} d^{c\dag}) \label{Eq:XcolorD}
\end{equation}
for $\mathcal{O}_{D}$.\footnote{ To be noticed, similar operators,
usually with two $X$'s, are used in common ADM models, where $X$ is
usually taken to be the DM particle.} In both cases, $X$ carries
baryon number $+1$, and zero lepton number. For
$\mathcal{O}_{Xq,S}$, $X$ is a $SU(2)_W$ singlet, and for
$\mathcal{O}_{Xq,D}$, $X$ is a $SU(2)_W$ doublet. One can introduce
multiple generations of $X$'s. This could lead to physical CP-
violating phase and further induce baryogenesis, i.e.
Hylogenesis.~\cite{Davoudiasl:2010am}.

To make the connection between DM particle and $X$, we introduce a
gauge singlet scalar field $\Phi_e$.  It couples to $X$ and DM
particles through the following Lagrangian:

\begin{equation}
 \mathcal{L}_{\Phi_e,S} =  v \overline{\phi}^2 \Phi_e^* + \lambda_s \Phi_e (X^c e^c) \label{Eq:PhieS}
\end{equation}
for $\mathcal{O}_{S}$, and

\begin{equation}
 \mathcal{L}_{\Phi_e,D} = v \phi^2 \Phi_e + \lambda_s \Phi_e^* (X L) \label{Eq:PhieD}
\end{equation}
for $\mathcal{O}_{D}$. $\Phi_e$ is a gauge single, and it carries
both baryon and lepton numbers as $+1$.  $v$ is a dimensionful
coupling in front of the 3-scalar operator.  We will discuss in
detail the suitable choices of various parameters in later content.

Now let us summarize the particle content in our model.  The DM
particle in our model is a scalar. It carries baryon and lepton
numbers as $-\frac{1}{2}$.  We further introduce $X$ and $\Phi_e$ to
link DM particle with SM particles. The properties of various
particles are summarized in Tab.~\ref{Eq:U1Assign}.
\begin{eqnarray}
 \begin{matrix}
 &&SU(3)_C&SU(2)_L&U(1)_Y
 &U(1)_L&U(1)_B \cr
 X &&{\bf 1 } &{\bf 1}({\bf 2})&1/2&0& 1 \cr
  \phi &&{\bf 1}& {\bf 1}& 0 &1/2&1/2\cr
 \Phi_e && {\bf 1}& {\bf 1 }&
 0 & 1 & 1\cr
 \end{matrix} \label{Eq:U1Assign}
\end{eqnarray}

After integrating out the heavy degree of freedom,\footnote{In this
study, we mainly focus on the DM induced nucleon decay processes.
Such processes distribute the nucleon mass, 1 GeV, to the final
state particles.  The typical 4-momentum in the internal propagators
are always smaller than 1 GeV.  Thus it is valid to integrate out
$\Phi_e$ as a heavy particle to generate the effective operator, as
long as its mass is larger than nucleon mass.} we are left with the
following effective operators:
\begin{equation}
 \mathcal{L}_{e,S} \supset \frac{\lambda_s v}{\Lambda^2 M_X M_{\Phi_e}^2} \overline{\phi}^2(e^c u^c)(d^c u^c) \label{Eq:EffLagS}
\end{equation}
for $\mathcal{O}_{S}$, and

\begin{equation}
 \mathcal{L}_{L,\mathrm{eff}} \supset \frac{\lambda_s v}{\Lambda^2 M_X M_{\Phi_e}^2} \overline{\phi}^2(L^\dag Q^\dag)(u^c d^c) \label{Eq:EffLagD}
\end{equation}
for $\mathcal{O}_{D}$.

Here we want to emphasize that although our effective operators are
dimension 8, only $\Lambda$ and $M_X$ have to be very large due to
collider constraints.  $\Phi_e$ is a gauge singlet. Its mass is
barely constrained. Moreover, we have a dimensionful parameter $v$ in
the numerator.  We will discuss the constraints for each parameter
carefully in the next section, and we will see that we could have a
sizable interaction cross section with a reasonable choice of
parameters.

\section{Choice of parameters}\label{SEC::Parameters}
In the previous section, we introduce the UV model to the effective
operators $\mathcal{O}_{S}$ and $\mathcal{O}_{D}$. In this section,
we will focus on the various constraints on parameters, and we
choose a benchmark point for the further study.

$\bullet$ DM mass:

As we have discussed in Sec.~\ref{SEC::Intro} and
Sec.~\ref{SEC::Lag}, we only focus on the model where DM carries
anti-baryon number.  If one starts with zero baryon and lepton
numbers, DM mass is naturally set to be 2-3 GeV. We will choose DM
mass as $3$ GeV as a benchmark point.  However, baryogenesis is not
our focus in this paper. Thus we do not constrain DM mass to be a
particular value for various signature studies.  DM mass cannot be
arbitrarily low.  To avoid nucleon decaying to two DM particles, DM
mass needs to be larger than 0.5 GeV.

$\bullet$ Dimensionful parameter $v$:

The dimensionful parameter $v$ can be in principle sizable. However,
large $v$ will induce large corrections to scalar masses through the
loop diagram.  Thus we require the loop corrections to the scalar
masses to be smaller than their bare values.  This implies
$v\lesssim 4\pi\ \mathrm{Min}\{m_{\mathrm{DM}},m_{\Phi_e}\}$.

$\bullet$ Mass of $\Phi_e$:

$\Phi_e$ is a gauge singlet. Thus the experimental constraints on
its mass is not very strong.  However, $\Phi_e$ should not be
lighter than $1$ GeV.  It carries both baryon and lepton number. If
its mass is smaller than $1$ GeV, then the proton can decay to
$\Phi_e$ and a positron.

The mass of $\Phi_e$ affects the picture of baryogenesis in our
model.  If $m_{\Phi_e}$ is larger than twice of $m_\phi$, then one
can directly applies the similar story as Hylogenesis, i.e. the
decay of $X$ and $\bar{X}$ induces the asymmetry for both $\Phi_e$
and SM sector through interference.  When $\Phi_e$ later decays to
$\phi$'s, the asymmetry in $\Phi_e$ is transferred into $\phi$.

If $\Phi_e$ is lighter than twice of $m_\phi$, asymmetry in $\Phi_e$
cannot propagate to $\phi$'s through its decay. Instead, $\Phi_e$
will decay to anti-proton and positron caused by the higher
dimensional operator.  This washes out the asymmetry in SM sector.
In this scenario, the asymmetry of DM must be deposited to $\phi$
through the off-shell $\Phi_e$ during $X$ decay.  The generated $\phi$
can annihilate with each other to anti-protons and positrons, i.e.
$\phi+\phi\rightarrow\bar{u}+\bar{d}+\bar{d}+e^+$.  This is the only
process which can wash out the asymmetry in $\phi$. However, as long as the
reheating temperature is lower than the freeze-out temperature of
such wash-out process, the asymmetry of $\phi$ is not removed.

As we discussed previously, the 3-scalar coupling, $v$, is chosen to
be large, i.e. $4\pi$ times the scalar mass scale.  Such a large
coupling compensates the phase space suppression in the 3-body decay
process.  Thus the 3-body decay branching ratio of $X$ is comparable
to its 2-body decay branching ratio.  Then one expects the asymmetry
generated when $m_{\Phi_e}<2m_\phi$ is comparable to the asymmetry
when $m_{\Phi_e}>2m_\phi$. In later discussion, we choose
$m_{\Phi_e}=m_\phi = 3$ GeV as our benchmark point.

One may worry whether the decay of $\Phi_e$ causes any problems of
BBN since its decay lifetime may be long by decaying through the
higher dimensional operator. In addition, the reheating temperature
needs to be high enough to induce BBN, whether that washes out the
asymmetry generated before is another concern.  We will cover these
subtleties at the end of this section.

$\bullet$ Mass of $X$ and cut-off scale $\Lambda$:

$X$ couples to SM particles through two vertices. $\lambda_s \Phi_e
(X^c e^c)$ or $\lambda_s \Phi_e^* (X L)$ links $X$ to leptonic
sector, and $\frac{1}{\Lambda^2} (X u^c)(d^c u^c)$ or
$\frac{1}{\Lambda^2} (X^c Q)(u^{c\dag} d^{c\dag}) $ links $X$ to
hadronic sector.

Let us first address on the constraints from leptonic sector.  Given
$\lambda_s \Phi_e (X^c e^c)$ or $\lambda_s \Phi_e^* (X L)$, the
strongest constraint comes from the LEP mono-photon
search~\cite{Fox:2011fx,Abdallah:2003np,Achard:2003tx,Abbiendi:2000hh,Heister:2002ut}.
By requiring photon energy larger than $10$ GeV, they constrain the
product of cross section and acceptance to be smaller than $0.1$ pb.
This can be reinterpreted into the constraint of our parameters. We
study the monophoton channel using MadGraph~\cite{Alwall:2011uj}. As
long as $\frac{M_X}{\lambda_s}$ is larger than 0.5 TeV, the model is
safe from the LEP constraint.

The other constraint on such operators is muon
$(g-2)_\mu$.\footnote{We thank the referee for pointing out this
important constraint.} Operator $\lambda_s \Phi_e (X^c e^c)$ and
$\lambda_s \Phi_e^* (X L)$ can induce large contributions to
$(g-2)_\mu$.  The corrections to $(g-2)_\mu$ for such models has
been calculated and it is summarized in \cite{Jegerlehner:2009ry}.
The leading contribution scales with the mass of particles as
\begin{eqnarray}
\delta a_\mu \simeq
\frac{\lambda_s^2}{4\pi^2}\frac{m_\mu^2}{m_{\Phi_e}^2}\frac{m_\mu}{m_X}
\label{Eq:gminus2}
\end{eqnarray}
Given the fact that $\Phi_e$ is only few GeV, the corrections to
$(g-2)_\mu$ can be significant.  If we take our benchmark point,
i.e. Eq.~(\ref{Eq:BMP}), we get $\delta a_\mu\sim 10^{-7}$, which is
safely ruled out by the current measurement \cite{Hanneke:2010au}.
However, the precise value of $(g-2)_\mu$ highly depends on the UV
model. Especially, the leading contribution under $m_X$ expansion is
different by a minus sign when the scalar is a real scalar or a
pseudo scalar. To make sure the leading contribution of $(g-2)_\mu$
vanishes precisely, it is crucial to have both $\lambda_s \Phi_e (X^c
e^c)$ and $\lambda_s \Phi_e^* (X L)$ operators in our model.  The
linear combination of these two operators in the Lagrangian should
be written as, (here we dropped all the kinematic and mass terms)
\begin{eqnarray}
\mathscr{L}\supset \lambda_s L X_L \Phi_e+\lambda_s e^{c
\dag}X_R^{c \dag}\Phi_e^*+h.c.\label{Eq:Lforgminus2}
\end{eqnarray}
Here everything is written in two-spinor convention. $L$ is the SM
lepton doublet, and $e^c$ is the right handed lepton.  $X_L$ and
$X_R^c$ are $SU(2)_W$ doublet and singlet. Given such Lagrangian,
$(g-2)_\mu$ correction comes in as the next leading order,
\begin{eqnarray}
\delta a_\mu \simeq
\frac{\lambda_s^2}{4\pi^2}\frac{m_\mu^2}{m_{\Phi_e}^2}\frac{m_\mu^2}{m_X^2}
\label{Eq:gminus2next}
\end{eqnarray}
For our benchmark point, the correction to $(g-2)_\mu$ is very
small, $\delta a_\mu\sim 10^{-11}$.\footnote{As a side point, one
can easily fit $(g-2)_\mu$ anomaly in our model by tuning
parameters.}  By changing the suppression scales of the
operators coupling $X_L$ or $X_R^c$ to quarks can decouple one of
$X$'s in the IND processes.  Thus we can and will still treat $\lambda_s
\Phi_e (X^c e^c)$ and $\lambda_s \Phi_e^* (X L)$ as two different
scenarios in our later discussion for simplicity.

Next we consider the constraint for $m_X$ and $\Lambda$ from the
hadronic side.  Here we do not try to $UV$ complete the operator,
instead we do a simple parton level analysis and make a conservative
choice of parameters.  For simplicity, we choose $M_X \sim \Lambda$.
A more careful study which includes effects of mediator's width has
been carried out in~\cite{Davoudiasl:2011fj}, we refer readers to
those papers for more details.

For $\frac{1}{\Lambda^2} (X u^c)(d^c u^c)$, $X$ is an $SU(2)_W$
singlet.  Once $X$ is produced through this effective operator, it
decays to a charged lepton and DM particles.  The signature is $1\
\textrm{jet}\ +\ l^{\pm} \ +\ MET$, where $MET$ is missing
transverse energy. For $\frac{1}{\Lambda^2} (X^c Q)(u^{c\dag}
d^{c\dag}) $, there is also a monojet channel, i.e. $1\ \textrm{jet}
\ +\ MET$.

Let us first focus on the monojet channel. This puts constraint on
$\frac{1}{\Lambda^2} (X^c Q)(u^{c\dag} d^{c\dag}) $, where $X$
decays to neutrinos half of the time. The most recent result on
monojet search is from~\cite{CMS:rwa}. With a MET cut at $350$
GeV\footnote{We also checked other values of MET cut, the conclusion
does not change.}, the statistical uncertainty of the monojet cross
section is about 4.5 fb.  If systematic uncertainties are included,
the error bar can only increase.  To see how well the monojet search
constrain our parameter space, we choose a benchmark point $M_X\sim
\Lambda\sim 1\ \textrm{TeV}$. Without accounting for the
reconstruction efficiency at detector level, only applying the MET
cut at parton level already reduces $\sigma\times A$ to $4$ fb. A
more detailed collider study will only bring $\sigma\times A\times
\epsilon$ of our signal lower.  Thus we choose $M_X\sim \Lambda\sim
1\ \textrm{TeV}$ as our conservative benchmark point for further
study. To accommodate the LEP constraint, we further choose
$\lambda_s$ as $2$ for our benchmark point.

Now we check the constraint from $1\ \textrm{jet}\ +\ l^{\pm} \ +\
MET$. There is no concrete search optimized for this particular
channel so far.  If there is b-quark in the final state, the
signature is similar to single top production.  However, if we take
$M_X\sim \Lambda\sim 1\ \textrm{TeV}$, the parton level cross
section for $p+p\rightarrow X+ \textrm{jet}$ is about $30\
\textrm{fb}$. This is the total cross section for all 3 generations
of leptons. Without further event selection and reconstruction, this
cross section has already been much smaller than the uncertainty for
single top production.

The other channel can be relevant is the $W'$
search.~\cite{ATLAS:2012fpa} However, this search is optimized for
s-channel production. Thus there are 2 kinematic cuts applied to the
event selection:
\begin{eqnarray}
0.4 < p_{T}^l/E_T^{\mathrm{miss}} <1.5 \nonumber   \\
|\Delta\phi_{l,\mathrm{miss}}-\pi| < 0.2\pi   \label{Eq:WpSearch}
\end{eqnarray}
Taking our benchmark value, i.e. $M_X\sim \Lambda\sim 1\
\textrm{TeV}$, the event selection efficiency for our signal after
these two cuts is about $61\%$ at parton level. Assuming flavor
universality, the parton level cross section for electron or muon
channel after the kinematic cuts is reduced to 2 fb.  This is
smaller than the constraints from $W'$ search in any mass region.
Further, the $W'$ search optimized the $M_T^{\mathrm{min}}$ cut with
respect to a particular model, which may not be the optimized cut
for our signal. Thus our benchmark point is also safe from this
search.

To summarize our choice of various parameters, we present our
benchmark point as the following:

\begin{eqnarray}
&m_{\phi} &= 3\ \textrm{GeV}, \nonumber   \\
&m_{\Phi_{e}} &= 3\ \textrm{GeV}, \nonumber   \\
&v &= 4\pi m_{\Phi_e} = 3 \times 4\pi\ \textrm{GeV}\\
&\lambda_s &= 2, \nonumber   \\
&m_X &= \Lambda =  1\ \textrm{TeV}. \nonumber    \label{Eq:BMP}
\end{eqnarray}

Later we will take this benchmark point and estimate signature
reaches for various channels.  Here we want to emphasize that our
benchmark point is a conservative one, and the signal strength could
be larger.

Before we end this section, let us demonstrate in a bit more detail some subtleties
about the thermal history of our model. First, the reheating
temperature cannot be too low for a successful BBN, while it cannot
be too high to wash-out the asymmetry through
$\phi+\phi\rightarrow\bar{u}+\bar{d}+\bar{d}+e^+$. We calculate the
freeze-out temperature for the wash-out process. The annihilation
cross section is estimated as
\begin{equation}
\sigma v\simeq \frac{(\lambda_s v)^2}{2048 \pi^5}(\frac{E^2}{M_X^2
\Lambda^4})\label{Eq:phiphiWash}
\end{equation}
where $E$ is the typical energy of the process. If we choose the
parameters as in the benchmark point, Eq.~(\ref{Eq:BMP}), the
reheating temperature needs to be smaller than $m_{\phi}$, i.e.
$T_{RH} \lesssim m_\phi\sim 3$ GeV.  Such low reheating temperature
is also requested in Hylogenesis models.

At last, we estimate the decay lifetime of $\Phi_e$.  $\Phi_e$
cannot decay too late, or else it messes up the successful
prediction of BBN.  The decay lifetime of $\Phi_e$ can be easily
estimated as
\begin{equation}
\Gamma_{\Phi_e}\sim \frac{\lambda_s^2 m_{\Phi_e}^7}{4096\pi^5 M_X^2
\Lambda^4}\label{Eq:PhieDecay}
\end{equation}
Taking the benchmark point in Eq.~(\ref{Eq:BMP}), we get the decay
lifetime around $10^{-4}$s.  Thus the decay of $\Phi_e$ is also
generically safe from BBN constraint.

\section{Relating to Chiral Lagrangian}\label{SEC::Chiral}

In the previous section, we introduced the effective operators for
DM interacting with SM particles at parton level. In this section,
we show how the parton level operators are related to nucleons and
mesons through chiral Lagrangian. By expanding the chiral Lagrangian
in powers of $p_{\mathrm{meson}}/(4\pi f)$, where $f \approx 139$
MeV , one can calculate the cross section for the following
processes: $p^+ + \phi \rightarrow e^+ + \overline{\phi}$, $n + \phi
\rightarrow \bar{\nu} + \overline{\phi}$ and $p^+ + \phi \rightarrow
\pi^0 + e^+ + \overline{\phi}$.  These three processes turn out to
be the most important processes in the signature searches. In this
section, we follow closely~\cite{Claudson:1981gh} for our
calculation.

For $\mathcal{O}_{S}$ and $\mathcal{O}_{D}$, the effective
Lagrangian after expanding in flavor basis can be written as
$\mathcal{L}_{\mathrm{int}} = \sum_i C_i O_i$, where $C_i$ are
dimension $(-4)$ constants related to Eq.~(\ref{Eq:EffLagS}) and
(\ref{Eq:EffLagD}). $O_i$ can be written as


\begin{align}
O_{S1} &= \epsilon_{\alpha\beta\gamma} \phi \phi (d^{\alpha}_R u^{\beta}_R)(u^{\gamma}_R e_R) \\
O_{S2} &= \epsilon_{\alpha\beta\gamma} \phi \phi (s_R^{\alpha}
u_R^{\beta})(u_R^{\gamma} e_R)
\end{align}

\begin{align}
O_{D1} &= \epsilon_{\alpha\beta\gamma} \phi \phi (d^{\alpha}_R u^{\beta}_R)(u^{\gamma}_L e_L -d_L^{\gamma} \nu_L) \\
O_{D2} &= \epsilon_{\alpha\beta\gamma} \phi \phi (s_R^{\alpha} u_R^{\beta})(u_L^{\gamma} e_L -d^{\gamma}_L \nu_L)\\
O_{D3} &= \epsilon_{\alpha\beta\gamma} \phi \phi (d_R^{\alpha}
u_R^{\beta})(s^{\gamma}_L \nu_L)
\end{align}
$\alpha,\beta,\gamma$ are color indices.  Here we do not include the
operators with two strange quarks since we do not consider final
states with two mesons.

The corresponding chiral Lagrangian for $\mathcal{O}_{S}$ and
$\mathcal{O}_{D}$ is

\begin{equation}
\mathcal{L}_{S,int} \supset C_{R1} \beta \mathrm{Tr} \left[O
\xi^{\dagger} (B_R e_R) \xi \phi \phi \right] + C_{R2} \beta
\mathrm{Tr} \left[\tilde{O} \xi^{\dagger} (B_R e_R) \xi \phi \phi
\right] \label{Eq:RH_CL}
\end{equation}

\begin{eqnarray}
\mathcal{L}_{D,int} &\supset& C_{L1} \alpha \mathrm{Tr} \left[O \xi (B_L e_L) \xi \phi \phi - O' \xi (B_L \nu_L) \xi \phi \phi \right] \nonumber\\
&+& C_{L2} \alpha \mathrm{Tr} \left[\tilde{O} \xi (B_L e_L) \xi \phi \phi -\tilde{O'} \xi (B_L \nu_L) \xi \phi \phi \right] + C_{L3} \alpha \mathrm{Tr} \left[\tilde{O''} \xi (B_L \nu_L) \xi \phi \phi \right]
\label{Eq:LH_CL}
\end{eqnarray}
where $C_{L,R1} = C_{L,R2} = C_{L_3} = \frac{1}{M_s^4}$. For the
benchmark point we chose in Eq.~(\ref{Eq:BMP}), the suppression
scale $M_s$ can be related to the parameters in the Lagrangian by

\begin{equation}
\frac{1}{M_s^4}= \frac{1}{\Lambda^2} \frac{1}{M_X} \lambda_s
\frac{v}{m_{\Phi_e}^2} = \frac{1}{(104 \mathrm{GeV})^4}
\end{equation}
$B_{L/R}$ is the baryon matrix operator, $\alpha = -0.015
\mathrm{GeV^3}$ and $\beta = 0.014 \mathrm{GeV^3}$~\cite{Aoki:1999tw} are the overall
constants and $\xi = \exp(iM/f)$ where M is the meson matrix
operator.

\begin{equation}
M = \left( \begin{array}{ccc} \frac{\eta}{\sqrt{6}} + \frac{\pi^0}{\sqrt{2}}& \pi^+ & K^+ \\
\pi^- & \frac{\eta}{\sqrt{6}} - \frac{\pi^0}{\sqrt{2}}  & K^0 \\
K^- & \bar{K}^0 & -\sqrt{\frac{2}{3}} \, \eta \end{array} \right) , \quad
B = \left( \begin{array}{ccc} \frac{\Lambda^0}{\sqrt{6}} + \frac{\Sigma^0}{\sqrt{2}}   & \Sigma^+ & p \\
\Sigma^- & \frac{\Lambda^0}{\sqrt{6}} - \frac{\Sigma^0}{\sqrt{2}} & n \\
\Xi^- & \Xi^0 & -\sqrt{\frac{2}{3}}\, \Lambda^0 \end{array} \right) .
\end{equation}

The operator $O$ and $\tilde{O}$ are defined in the same way as in~\cite{Claudson:1981gh}

\begin{eqnarray}
O &=& \phantom{-}\left( \begin{array}{ccc} 0 & 0& 0 \\
0 & 0  & 0 \\
1 & 0 & 0 \end{array} \right) , \quad
O' = \phantom{-} \left( \begin{array}{ccc} 0 & 0 & 0 \\
0 & 0 & 0 \\
0 & 1 & 0 \end{array} \right),  \quad
\phantom{O' = \left( \begin{array}{ccc} 0 & 0 & 0 \\
0 & 0 & 0 \\
0 & 1 & 0 \end{array} \right),}\nonumber\\
\tilde{O} &=& -\left( \begin{array}{ccc} 0 & 0& 0 \\
1 & 0  & 0 \\
0 & 0 & 0 \end{array} \right) , \quad
\tilde{O'} = -\left( \begin{array}{ccc} 0 & 0 & 0 \\
0 & 1 & 0 \\
0 & 0 & 0 \end{array} \right), \quad
\tilde{O''} = \left( \begin{array}{ccc} 0 & 0& 0 \\
0 & 0  & 0 \\
0 & 0 & 1 \end{array} \right).
\end{eqnarray}

Now we expand the chiral Lagrangian to leading and next-to leading
order, which corresponds to no meson and a single meson respectively
in the final state.

For $\mathcal{O}_{S}$, the expansion of chiral Lagrangian is given
by
\begin{equation}
\mathcal{L}_\mathrm{int} \supset C_{R1} \beta p_R e_R \phi \phi - i \frac{C_{R1} \beta}{\sqrt{2}f_{\pi}}\pi^0 p_R e_R \phi \phi
\end{equation}\label{Eq:LRight}
and for $\mathcal{O}_{D}$,
\begin{equation}
\mathcal{L}_\mathrm{int} \supset C_{L1} \alpha p_L e_L \phi \phi + i \frac{C_{R1} \alpha}{\sqrt{2}f_{\pi}}\pi^0 p_L e_L \phi \phi
\end{equation}\label{Eq:LLeft}

In this paper, for the single meson channel, we only focus on the
pion channel, i.e. $p + \phi \rightarrow \pi^0 + e^+ +
\overline{\phi}$. This channel turns out to be the best search
channel for SuperK in our later study, see Sec.~\ref{SEC::IND}.

\begin{figure}[htb]
\centering
\includegraphics[width=0.5\columnwidth]{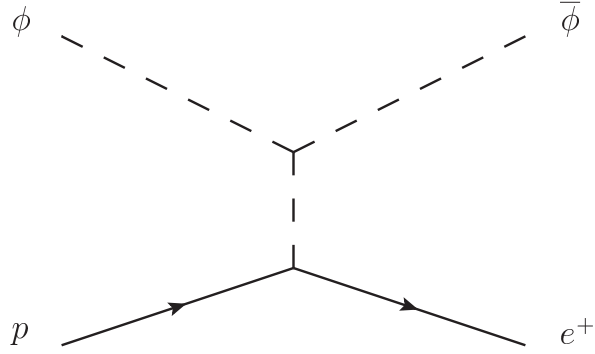}
\caption{Feynman Diagrams for $p + \phi \rightarrow e^+ + \overline{\phi}$ process in Chiral Perturbation Theory}\label{Fig:Chiral_2to2}
\end{figure}

\begin{figure}[htb]
\centering
\includegraphics[width=0.43\columnwidth]{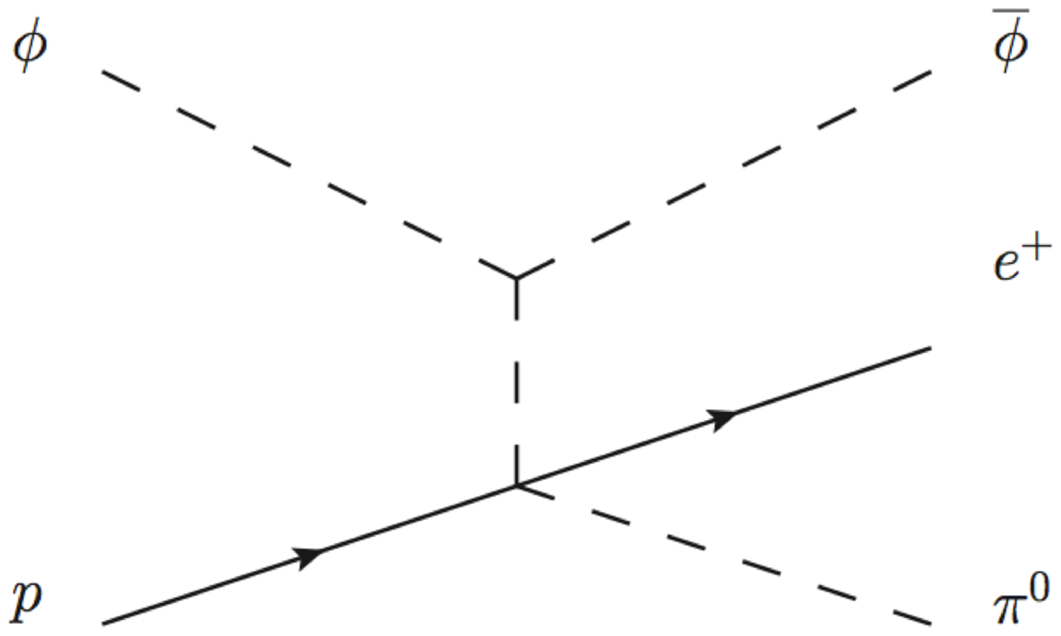}
\includegraphics[width=0.4\columnwidth]{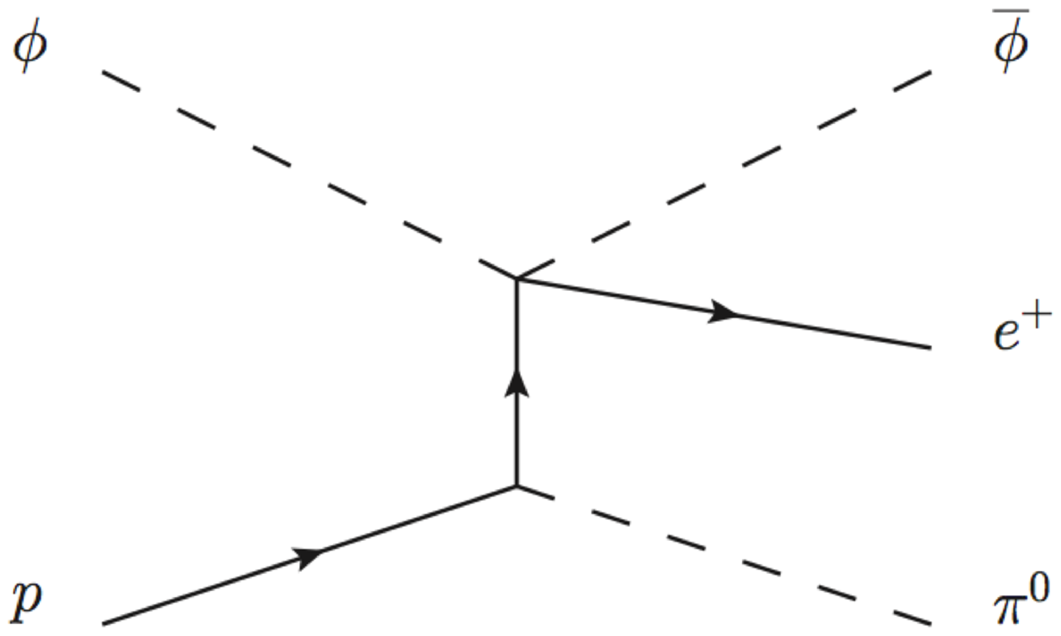}
\caption{Feynman Diagrams for $p + \phi \rightarrow \pi^0 + e^+ + \overline{\phi}$ process in Chiral Perturbation Theory}\label{Fig:Chiral_2to3}
\end{figure}

Here we want to emphasize that the leading processes in our model
are 2-to-2 processes, shown as Fig.~\ref{Fig:Chiral_2to2}.  The only
SM particles in the final state is charged lepton or neutrino.  Such
2-to-2 processes have much larger cross section comparing to 2-to-3
process.  Meanwhile, the 2-to-3 process has more visible particles
in the final states, as Fig.~\ref{Fig:Chiral_2to3}. This helps to
reconstruct the event. These two channels will lead to interesting
signatures to look for respectively.

To get an intuition for the interaction rate for each process, we
show the cross section for each process with parameters as our
benchmark point, i.e. Eq.~\ref{Eq:BMP}. The cross section for
2-to-2 process is
\begin{equation}
\sigma_{2-\mathrm{to}-2} = 1.87 \times 10^{-43} \mathrm{cm^2}
\end{equation}
And the cross section of the 2-to-3 process is
\begin{equation}
\sigma_{2-\mathrm{to}-3} = 2.36 \times 10^{-48}\mathrm{cm^2}.
\end{equation}
Given the estimations on the cross sections, in the following
section, we will focus on various signatures induced by this model,
and we will see how each search channel probes the parameter space.



\section{Signature searches}\label{SEC::signature}
DM particles annihilating with nucleons could induce several
interesting signatures different from ordinary DM search channels.
In this section, we focus on the experimental signatures of this
model.

First, we will study the signature in proton decay experiment, e.g.
Super-Kamiokande, in Sec \ref{SEC::IND}. Induced proton decay
process has been discussed in the context of magnetic monopoles
in~\cite{Dimopoulos:1982fj,Dimopoulos:1982cz}. Similar signatures
induced by asymmetric dark matter has also been considered
in~\cite{Davoudiasl:2010am,Davoudiasl:2011fj,Blinov:2012hq}.
However, we emphasize that similar signatures in our model have
different SM particles in the final states.  We benefit from having
our signal in the best search channel in proton decay experiments.
Also we study in detail on the event selection in such channel,
which may further help to improve the search capability.

Furthermore, if DM is captured by the Sun, it can annihilate with
the nucleon and may induce a large flux of anti-neutrinos. The
neutrino experiments could put strong constraints on such scenario.
At last, the anti-DM in the final state is boosted to a high
velocity, and it can escape the Sun.  Underground experiments may
also be able to detect such anti-DM flux from the Sun.  We will
address each of these signatures in this section.

\subsection{Induced Proton Decay in Super-Kamiokande}\label{SEC::IND}
DM particles in the cosmic background can interact with nucleon in
the proton decay experiments and induce signals similar to nucleon
decay.  Currently, the best nucleon decay experiment is
Super-Kamiokande~\cite{Nishino:2012ipa}, which puts stringent
constraints to various nucleon decay channels.  In this section, we
reinterpret the nucleon decay lifetime limit as a constraint on the
DM-nucleon interaction cross section, and study how that constrains
the parameter space in our model.

As we have seen in Sec.~\ref{SEC::Chiral}, the dominant annihilation
channel between DM and nucleon is through the 2-to-2 process, i.e.
$\phi+p^+ \rightarrow \overline{\phi}+ e^+$ or $\phi+ n
\rightarrow \overline{\phi}+ \bar{\nu}$.  However, such two channels
suffer from the large atmospheric neutrino background, which could
interact with nucleon through either charge-current or
neutral-current interaction.  We are forced to consider the next
leading processes where one meson is included in the final state.

The most constrained channel in nucleon decay experiments is $p^+
\rightarrow e^+ + \pi^0$.  Since we have $\phi+p^+ \rightarrow
\overline{\phi}+ e^+ +\pi^0$ in the DM-nucleon annihilation process,
this channel shares the same visible final states with the best
proton decay channel, we will focus on this process and study the
implication of the current decay lifetime constraint.

For each proton, the effective decay lifetime can be calculated as
the inverse of the interaction rate:

\begin{eqnarray}
\tau_{\mathrm{eff}}= \frac{1}{n_{\mathrm{DM}} (\sigma v)_{\mathrm{IND}}}
\label{Eq:Lifetime}
\end{eqnarray}

We take the DM energy density around the Earth as $0.3
\frac{\textrm{GeV}}{\textrm{cm}^3}$.  The annihilation cross section
between DM particle and proton for the benchmark point in
Eq.~(\ref{Eq:BMP}) is $0.707 \times 10^{-40} \textrm{cm}^3/s$. Thus
the effective proton decay lifetime is

\begin{eqnarray}
\tau_{\mathrm{eff}}= 1.5\times 10^{33} \textrm{yr} (\frac{0.7\times 10^{-40}
\mathrm{cm^3/s}}{(\sigma v)_{\mathrm{IND}}})\label{Eq:LifetimeBMP}
\end{eqnarray}

A proton lifetime of $1.5\times 10^{33} ~\textrm{yr}$ is shorter
compared to the current experimental bound from SuperK for proton
decay in this channel ($\tau_p = 8 \times 10^{33}
~\textrm{yr}$)~\cite{Nishino:2012ipa}. However, this is before any
event selection and reconstruction efficiencies are considered.
Since our process is a 2-to-3 scattering process, the kinematics are
different from real proton decay. The difference in kinematic
distributions can lead to different event selection efficiency. This
will further affect the interpretation from the proton decay
lifetime to the interaction rate in our model.

For our process, the final state ($\pi^0$ and $e^+$) reconstruction
efficiency is the same as proton decay process, since we share the
same final state particles within similar energy region.  To get rid
of the large background from atmospheric neutrinos, SuperK further
requests the following two event selection
cuts~\cite{Nishino:2012ipa}:

$\bullet$ The reconstructed proton's momentum, $p_P$, needs to be
smaller than $250\ \textrm{MeV}/c$.

$\bullet$ The reconstructed proton's invariant mass, $M_{\mathrm{INV}}$,
needs to be between $800\ \textrm{MeV}/c^2$ and $1050\
\textrm{MeV}/c^2$.

\begin{figure}[htb]
\centering
\includegraphics[width=0.45\columnwidth]{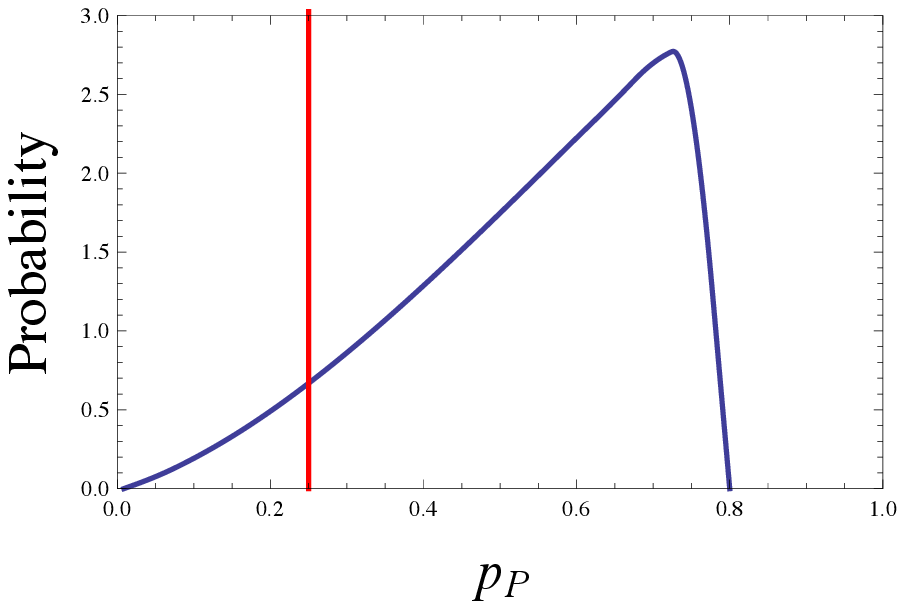}
\includegraphics[width=0.45\columnwidth]{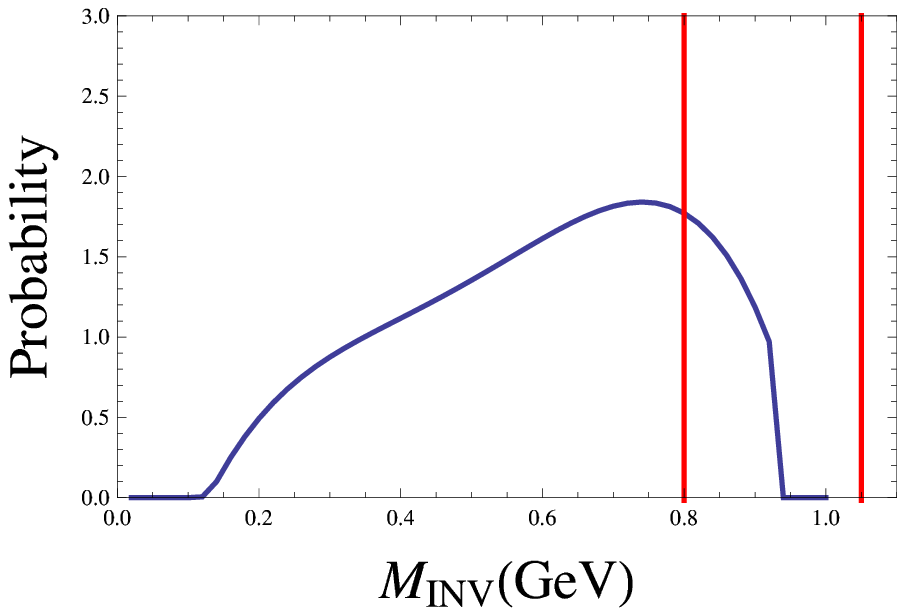}
\caption{Reconstructed proton transverse momentum and invariant mass
cut effeciency of induced nucleon decay searches in SuperK. Red
lines in the plots indicate the event selection cuts applied by
SuperK.}\label{Fig:Cut}
\end{figure}

To estimate the event selection efficiency of these two constraints,
we do a MC simulation of our process, as shown in
Fig.~\ref{Fig:Cut}. One can see that the selection efficiency of
these two cuts is very low. Only $5.23\%$ of signal events pass the
cuts.  Taking into account of both final state reconstruction
efficiency and event selection efficiency, we get an effective
proton decay lifetime of $\tau\simeq 2.9 \times 10^{34} yr$ for our
benchmark point in Eq.~(\ref{Eq:BMP}). Such a proton decay lifetime
could be reached by Hyper-Kamiokande around 2023.~\cite{Abe:2011ts}
One possibility to improve the experimental reach is to loosen the
event selection cuts in a reasonable way.  In SuperK analysis, they
provide detailed distributions of signal and background on the
$M_{\mathrm{INV}}-p_P$ plane~\cite{Nishino:2012ipa}.  If one loosens
the $p_P$ cut from $250 \mathrm{MeV}/c$ to $400 \mathrm{MeV}/c$, the
atmospheric neutrino background are still almost completely removed.
However, loosening the cuts in such a mild way can dramatically
increase our signal selection efficiency from $5.23\%$ to $20\%$.
This brings the effective proton decay lifetime to $7.5\times
10^{33} \mathrm{yr}$. Such decay lifetime could have already been
probed by SuperK in 2007.\footnote{Here we want to emphasize that
since we loosen the $p_P$ cut of the event selection, one cannot use
the current SuperK reach estimation for a reliable interpretation.
To optimize the cuts respect to our signal, a careful study is
necessary. However, this is out of the scope of this paper. Here we
provide a naive estimation on the reach capability. This provides an
intuition on how much better one can probe the parameter space by
optimizing the cuts for the DM induced proton decay process. }

\subsection{Signatures from the Sun}\label{SEC::Capture}
In the previous section, we consider the DM particles in the cosmic
background interact with the protons in SuperK.  In this section, we
consider the possible signatures induced by DM captured in stellar
objects.  Being asymmetric, DM cannot annihilate with each other. A
large number density of DM can exist in the stellar objects.
Further, the nucleon number density of the stellar object is usually
much higher than matter on the Earth, one expects the IND process
happens much more frequently.  Here we want to emphasize that
in~\cite{McDermott:2011jp,Goldman:1989nd,Gould:1989gw,Bertoni:2013bsa},
one constrains the scalar asymmetric DM models by requiring that the
accumulation of DM in neutron star does not cause a black hole in
the core.  In our model, since DM can annihilate with nucleons, and
the anit-DM in the final state of IND process can further annihilate
with DM, the accumulation of DM in the neutron star is not efficient
enough to form a black hole. Thus the bound is evaded.

Since the Sun has both large DM capture rate and relatively short
distance to the Earth, it provides the best place to look for
signatures.  We will focus on the Sun in the following discussion.
For various final states in IND processes, mesons and charged
leptons cannot propagate out. To observe such processes on the
Earth, we have to rely on the weakly interacting particles, i.e.
anti-neutrinos or anti-DM. We first set up the calculation of
capture rate and IND interaction rate. Then we will focus on the
possible signatures from the anti-neutrino flux in
Sec.~\ref{SEC::Neutrino} and anti-DM flux in
Sec.~\ref{SEC::anti-DM}.

\subsubsection{Dark matter accumulation in the Sun}
In this section, we calculate the accumulation of DM particles in
the Sun.\footnote{A general discussion of the process can be found
in Appendix.~\ref{SEC::GENERAL}.  We refer the reader to
~\cite{Griest:1986yu,Gould:1991hx,Hooper:2008cf} for details.}
Instead of the IND process, a dark matter particle can also
elastically scatter with the hydrogen and helium of the Sun and
become captured. The capture rate has been studied
by~\cite{Gould:1991hx}. In Appendix~\ref{SEC::GENERAL}, we provide a
general formula for capture rate. In the range of few GeV DM mass,
we can approximate the capture rate to be:

\begin{equation}
C^\odot \simeq 1.3 \times10^{25}\mathrm{s}^{-1}\left(\frac{\rho_{\mathrm{DM}}}{0.3\mathrm{
GeV/cm^3}}\right)\left(\frac{270\mathrm{km/s}}{\bar{v}}\right)\left(\frac{1\mathrm{GeV}}{m_{\mathrm{DM}}}\right)\left(\frac{\sigma_{\mathrm{elas}}}{10^{-40}\mathrm{cm^2}}\right).
\label{Eq:CaptureSun}
\end{equation}

For dark matter mass larger than 10 GeV, an additional kinematic
suppression factor needs to be
applied.\footnote{Eq.~(\ref{Eq:CaptureSun}) provides a general
feeling of the capture rate dependance on the main parameters. The
calculations carried in the paper is based on the more accurate
equations in Appendix~\ref{SEC::GENERAL}.} For light DM, the elastic
scattering cross section is not strongly constrained by current
direct detection experiments.  Thus for light DM mass region, we
take $\sigma_{\mathrm{elas}}$ to be $10^{-39} \mathrm{cm^2}$ for
spin-independent cross section and $10^{-36} \mathrm{cm^2}$ for
spin-dependent cross section. For large DM mass, we assume the
elastic scattering cross section to be the largest value allowed by
various direct detection searches.

Dark matter particle can thermalize with the Sun after being
captured, if DM mass is light, the evaporation from the Sun is not
negligible.  According to~\cite{Griest:1986yu}, the
evaporation rate can be estimated as

\begin{equation}
E^\odot \simeq
10^{-\left(\frac{7}{2}(m_{\mathrm{DM}}/\mathrm{GeV})+4\right)}s^{-1}\left(\frac{\sigma_{\mathrm{elas}}}{5\times10^{-39}\mathrm{cm^2}}\right)
 \label{Eq:EscapeRate}
\end{equation}

\begin{figure}[htb]
\begin{center}
\hspace*{-0.75cm}
\includegraphics[width=0.6\textwidth]{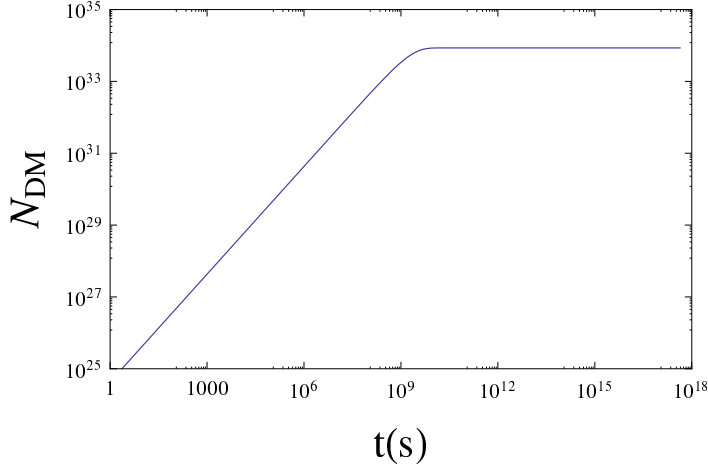} \hspace{1cm}
\caption{An illustration of the ADM number evolution in the Sun. We
choose the parameters according to the benchmark point where the IND
cross section is $5.6\times 10^{-36}\mathrm{cm^3/s}$. And we take the
elastic scattering between DM and nucleons as $10^{-40}\mathrm{cm^2}$. One
see that the number of ADM approaches a constant at late time due to
the equilibrium between capture and IND annihilation.}
\label{fig:ADMEvolve}
\end{center}
\end{figure}

Thus one can write the evolution equation for the DM captured in the
Sun as
\begin{eqnarray}
\frac{\mathrm{d} N_{\mathrm{DM}}}{\mathrm{d}t}=C^{\odot} - N_{\mathrm{Flavor},i}(\sigma v)_{\mathrm{IND}} (\rho_{c,i}/m_i)
N_{\mathrm{DM}}-E^{\odot} N_{\mathrm{DM}}, \label{Eq:Evolution}
\end{eqnarray}
where $(\sigma v)_{\mathrm{IND}}$ is the IND interaction cross
section. Unlike in the proton decay search in SuperK, we are looking
for the final states of the IND process such as anti-neutrino and
anti-DM fluxes. The IND process is dominated by the 2-to-2
scattering channels, i.e. $\phi+p^+ \rightarrow \overline{\phi} +
e^+$ and $\phi + n \rightarrow \overline{\phi} + \bar{\nu}$.  For
our benchmark point, i.e. Eq.~(\ref{Eq:BMP}), we get $(\sigma
v)_{\mathrm{IND}}$ for the 2-to-2 process as $5.6\times 10^{-36}
\mathrm{cm^3/s}$. This is a much larger interaction cross section
comparing to the 2-to-3 process in SuperK search. $i= n,p$. For
$\mathcal{O}_S$, $N_{\mathrm{Flavor,n}} = 0$ and
$N_{\mathrm{Flavor,p}} = 2$. For $\mathcal{O}_D$,
$N_{\mathrm{Flavor,n}} = 3$ and $N_{\mathrm{Flavor,p}} = 2$. Since
$\mathcal{O}_S$ cannot generate anti-neutrinos in the final state at
leading order, we will only focus on $\mathcal{O}_D$ when we discuss
anti-neutrino flux in Sec.~\ref{SEC::Neutrino}. On the other hand,
both $\mathcal{O}_S$ and $\mathcal{O}_D$ can generate anti-DM flux.
In Sec.~\ref{SEC::anti-DM}, we also use $\mathcal{O}_D$ for
illustration.\footnote{The result from $\mathcal{O}_S$ is only
different by an $O(1)$ factor.} In Eq.~(\ref{Eq:Evolution}), we do
not include the DM pair annihilation term since the anti-DM produced
through IND processes in the Sun will escape the Sun after
production, as will be discussed in detail in
Sec.~\ref{SEC::anti-DM}.  $\rho_{c,i}$ is the mass density of
protons and neutrons in the center of the Sun and $m_i$ is the
proton and neutron masses. Here we show an illustration on the
evolution of ADM number in the Sun for our choice of benchmark point
as Fig.~\ref{fig:ADMEvolve}.  We see that the number of ADM
approaches a constant at late time due to the equilibrium between
capture and IND annihilation.

\subsubsection{Anti-Neutrino flux from the Sun}\label{SEC::Neutrino}
In this section, we focus on the anti-neutrino flux induced by IND
process in the Sun.  For IND process, DM annihilates with neutrons
producing anti-neutrinos in the final state. Neutrons are mainly
from helium, which is about $28\%$ mass of the Sun. Since the
kinematic energy of the DM particle at the core of the Sun is much
smaller than 1 GeV, one can treat the DM particles as at rest for
approximation, and then the anti-neutrino from the DM-nucleon
annihilation is monochromatic,
\begin{eqnarray}
p_{\bar{\nu}} = \frac{2 m_{\mathrm{DM}}+m_{\mathrm{N}}}{2(m_{\mathrm{DM}}+m_{\mathrm{N}})}m_{\mathrm{N}}
\label{Eq:NeutrinoEnergy}
\end{eqnarray}
For example, if DM particle mass is 3 GeV, the neutrino in the final
state is about 0.88 GeV.

The flux of atmospheric neutrinos has been measured by FREJUS
Collaboration ~\cite{Rhode:1996es}, the result agrees with the
theoretical calculation~\cite{Honda:2006qj,Honda:2011nf}. Since the
neutrino from IND process is monochromatic, we only need to focus on
one energy bin of the spectrum.  For example, for 0.88 GeV neutrino,
the corresponding bin is from 0.76 GeV to 1.00 GeV in FREJUS.  The
dominant uncertainty is coming from the theoretical uncertainty of
the interaction cross section between neutrino and nucleon.
Combining all the uncertainties, the error of the bins around 1 GeV
is about $22\%$. By requiring the contribution to neutrino flux from
IND process to not exceed 2 sigma error bar, one can constrain the
interaction rate of IND process in the Sun.

Note that one can further probe the parameter space of our model by
optimizing the neutrino flux measurement.  The atmospheric neutrino
flux measurement, i.e.~\cite{Rhode:1996es}, does not include the
angular information of the neutrino.  Since anti-neutrinos from IND
process is dominantly from the center of the Sun, the angular
information can help to reduce the atmospheric neutrino background
dramatically.  Furthermore, since the anti-neutrinos from the IND
process is monochromatic, a better energy resolution also helps to
improve the signal reach.  To get an idea on how these improvements
may help us probe the parameter space, we quote the energy and
angular resolution from the proposal of
ICANOE\footnote{The ICANOE proposal is based on technology in 1999. With current technology, the resolution may have been improved}~\cite{Cavanna:1999yj}. Since ICANOE is using the information
of all particles in the final state, it can achieve a good
reconstruction of incoming neutrinos' energy and incidence angle.
For the neutrino flux spectrum, energy resolution in ICANOE can be
as good as 50 MeV, this is about a factor of 5 improvement comparing
to the energy resolution of FREJUS Detector.  For neutrinos at
around 1 GeV, the angular resolution of the incoming neutrino is
about 12 degrees. With the angular information, the background can
be reduced by a factor of 90.  If one fully applies both energy and
angular resolution, the number of background events can be reduced
dramatically. However, given an exposure of 50 $kton\times year$,
there will be about 15 events in each bin of fixed energy and
incident angle.  The statistical uncertainty becomes comparable to
the theoretical uncertainty. Thus we take conservative choices of
energy and angular resolutions, assuming they can reduce the number
of background events by a factor of 200.

\begin{figure}[htb]
\begin{center}
\hspace*{-0.75cm}
\includegraphics[width=0.50\textwidth]{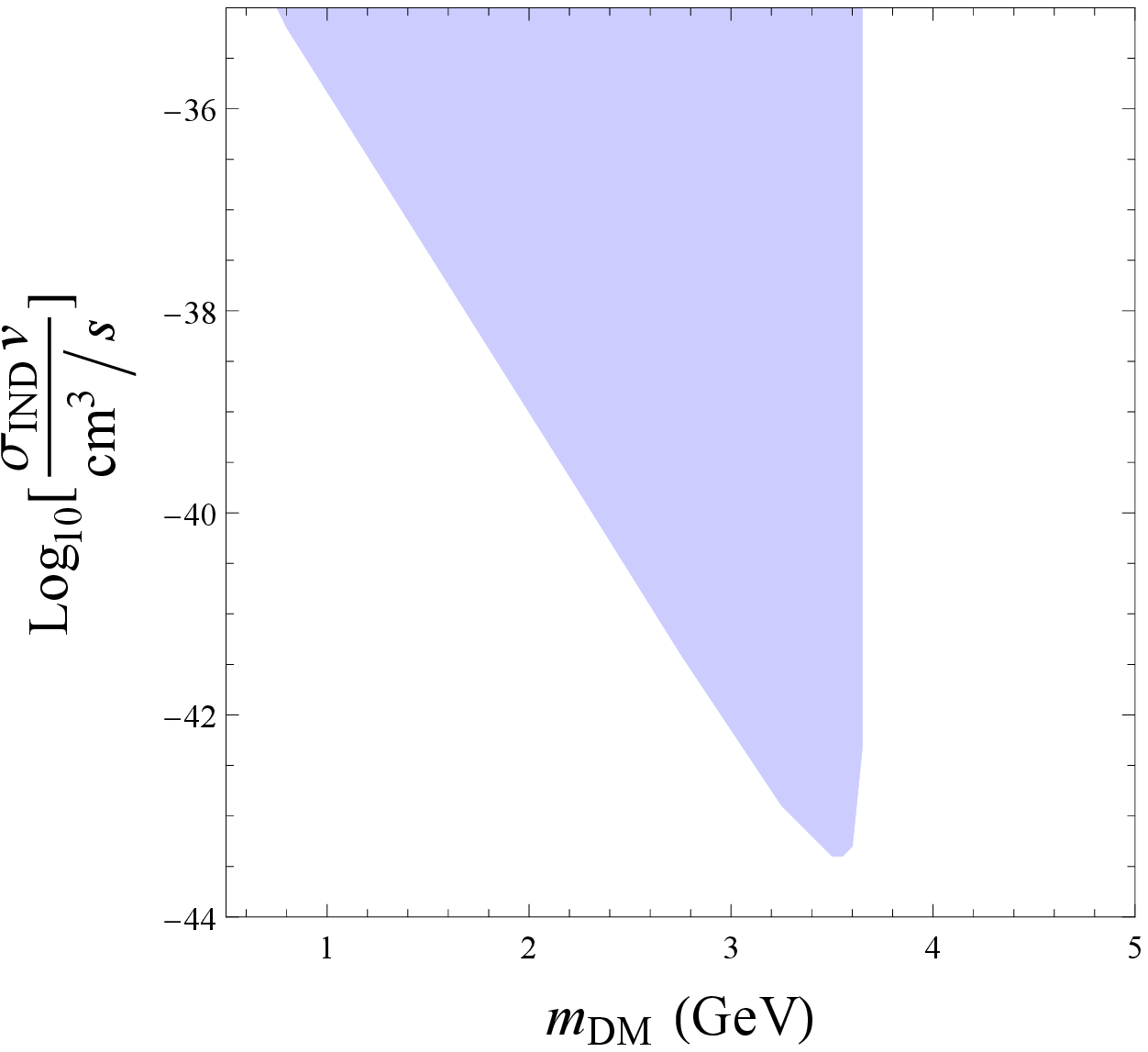} \hspace{0.1cm}
\includegraphics[width=0.50\textwidth]{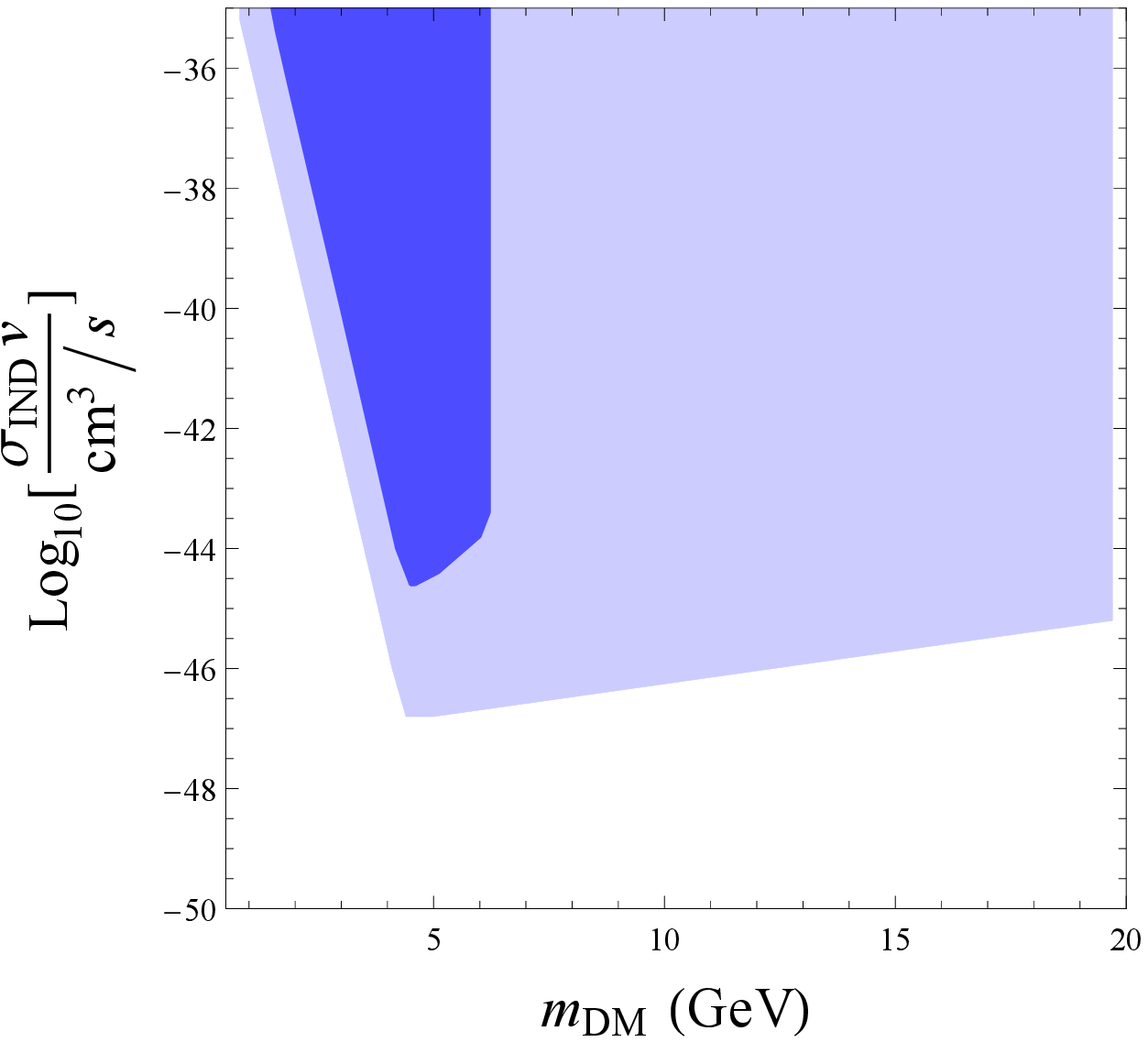}
\caption{The constraint from neutrino flux on the IND interaction
cross section as a function of the DM mass, for Spin-Independent
(Left) and Spin-Dependent (Right) elastic scattering respectively.
The dark blue region is constrained by the current data.  The light
blue is the region could be further probed by improving energy
resolution and angular information. For spin-independent scattering,
the current data of neutrino flux cannot probe any interesting
parameter space due to the low capture rate.  Taking other
parameters the same as our benchmark point, the lowest points in
both plots can be interpreted as the cut-off scales, i.e. $m_X\sim
\Lambda $. For spin-independent scenario, $\Lambda_{\mathrm{SI,Max}} \sim 24~\mathrm{TeV}$, and $\Lambda_{\mathrm{SD,Max}} \sim 91~\mathrm{TeV}$ for spin-dependent
scenario. } \label{fig:NeutrinoSun}
\end{center}
\end{figure}

To calculate the IND process rate in the Sun, one needs to specify
the elastic scattering cross section.  This is constrained by
various direct detection experiments.  For spin independent
scattering, when DM mass is larger than 5.5 GeV, the strongest
constraint comes from the recent LUX result~\cite{Akerib:2013tjd}.
Between 3.5 GeV to 5.5 GeV, the CDMS-Lite~\cite{Agnese:2013lua} sets
the best constraints.  Below 3.5 GeV, there is no constraint. (See
~\cite{Kusenko:2013saa} for more information.) For the spin
dependent case, since the Sun is dominated by protons, we focus on
the direct detection constraints for DM-proton elastic scattering.
The constraints dominantly come from PICASSO, SIMPLE and
COUPP~\cite{Archambault:2012pm,Felizardo:2010mi,Behnke:2012ys}. If
DM mass is smaller than 4 GeV, the constraints are not strong. To
estimate how well the anti-neutrino flux can help to probe our
parameter space, we assume the elastic scattering cross section to
be just below the constraints from various experiments.  When DM
mass is too small to be constrained, we take $10^{-39} \
\textrm{cm}^2$ and $10^{-36} \ \textrm{cm}^2$ for spin-independent
and spin-dependent cross sections respectively.  In Fig.~
\ref{fig:NeutrinoSun}, we show the
$(\sigma_{\mathrm{IND}}-m_{\mathrm{DM}})$ plane which can be probed
by the anti-neutrino flux. The dark blue is the region which is
constrained by the current data of atmospheric neutrino flux
measurement, assuming the largest elastic scattering cross section
allowed by direct detections. The light blue is the region which
could be constrained using a better energy resolution and angular
information.

From the plot, one can clearly see that the escape rate starts to
dominate the loss of the DM particles in the Sun when DM mass is
smaller than 4 GeV.  If DM particle scatters with nucleon
spin-independently, the current neutrino flux measurement cannot
probe any interesting parameter space due to the low capture rate.
However, if we apply a better energy resolution and angular
information as claimed in ICANOE, an interesting region can be
probed. The smallest IND cross section can be probed in this
scenario is about $3 \times 10^{-43}\mathrm{cm^3/s}$, which is corresponding
to $m_X\sim \Lambda \sim 24\ \mathrm{TeV}$, assuming all other
parameters to be the same as our benchmark point. Further, if the
elastic scattering is spin-dependent, the capture rate is much
higher, and one can probe a much larger parameter region, both
smaller IND cross section and larger DM mass. Using the current
measurement, the smallest IND cross section can be probed is about
$10^{-45}\mathrm{cm^3/s}$, which corresponds to $m_X\sim \Lambda
\sim 42\ \mathrm{TeV}$.  If we apply the improvements on energy
resolution and angular information, one can reach $m_X\sim \Lambda
\sim 91\ \mathrm{TeV}$.

\begin{figure}[t]
\begin{center}
\hspace*{-0.75cm}
\includegraphics[width=0.5\textwidth]{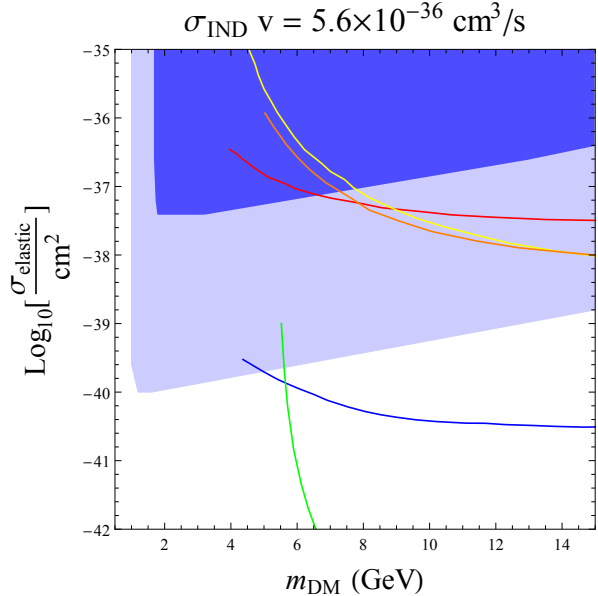} \hspace{1cm}
\caption{Assuming the IND cross section as our benchmark point, one
can constrain the elastic scattering cross section between DM with
nucleon.  The solid lines are indicating the constraints from direct
detections for spin-dependent and spin-independent scattering.  The
red curve is from PICCASO, orange is from SIMPLE, yellow is from
COUPP, Blue is from CDMSLite and Green is from
LUX.~\cite{Archambault:2012pm,Felizardo:2010mi,Behnke:2012ys,Akerib:2013tjd,Agnese:2013lua}}
\label{fig:ElasXsecNeuSun}
\end{center}
\end{figure}

On the other hand, if one takes a particular value of the IND cross
section, one can constrain the elastic scattering cross section. For
example, if we take the benchmark point where the IND cross section
is  $5.6\times 10^{-36}\mathrm{cm^3/s}$, the constraint on elastic scattering
cross section is show as Fig.~\ref{fig:ElasXsecNeuSun}. The
constraint for spin-dependent cross section is much stronger than
direct detection, while a stronger constraint can only be applied in
the lower mass region for spin-independent elastic scattering.

\subsubsection{Anti-DM flux from the Sun}\label{SEC::anti-DM}
In this section, we discuss another signature in our model, i.e. the
anti-DM flux from the Sun. Anti-DM is in the final state of the
2-to-2 induced nucleon decay process, its momentum can be calculated
easily by assuming that the initial particles are approximately at rest,
\begin{eqnarray}
p_{\overline{\phi}}=p_{l} = \frac{2 m_{\mathrm{DM}}+m_{\mathrm{N}}}{2(m_{\mathrm{DM}}+m_{\mathrm{N})}}m_{\mathrm{N}}
\end{eqnarray}
For example, if the DM mass is 3 GeV, then the velocity of anti-DM
in the final state is about $0.3\ c$.  This is much larger than the
escape velocity.  Thus the IND process can generate an anti-DM flux
from the Sun.  When arriving at the Earth, these fast anti-DM
particles can elastically scatter with the nucleus in underground
experiment detectors. Due to the large velocity of the anti-DM, it
can kick the neutron or proton out of the nucleus. A fast
neutron/proton plus a prompt gamma ray from the nucleus
de-excitation is the signature of the anti-DM flux.

For a 3 GeV DM particle, the typical velocity of the neutron/proton
after the elastic scattering is about $0.4\ c\sim 0.5\ c$. In
SuperK, such proton is not fast enough to generate the Cherenkov
ring. \footnote{Here we note that semi-annihilation DM models
\cite{D'Eramo:2010ep} may also generate fast moving DM
flux from the Sun.  Therefore, the search proposed in this
section can also be applied. The velocity of the DM particle in the
final state is model dependent, but generically higher than that of
IND process. We will leave the discussion of semi-annihilation
models in the future.} Meanwhile, the fast moving neutron can be
captured by the hydrogen and release a 2.2 MeV gamma ray. The
efficiency for SuperK to see such low energy gamma ray is low, only
about $20\%$. However, if one dopes Gd ion into the water, which is
being tested by SuperK, the fast neutron can be captured and
releases a gamma ray at about 8 MeV. \footnote{We gratefully thank
Michael Smy and Henry Sobel for very helpful discussions on details
about SuperK.}  This could help in triggering our signal.
Furthermore, the fast moving proton can leave a long track in the
detector since the stopping power is only about $O(1)$ MeV/cm for
few hundred MeV proton~\cite{Berger,Agostinelli:2002hh}. A Gd dopped
liquid scintillator detector, e.g. in Daya Bay
experiment~\cite{Guo:2007ug}, though much lighter than SuperK, can
provide much more information about the event, such as the incidence
energy and angle. This can help to reduce the background
efficiently. A detailed study on the experimental details and how to
optimize the signal are necessary for the search of this signature,
we will leave the details for future study.

A fast-moving proton leaves a long track in material, which is a
promise signal on which to trigger.  Also there are possibilities to
improve the signal with energy and angular information from the
proton track, we focus on the signature where the anti-DM knocks out
proton from the oxygen nucleus. For current study. we look for the
anti-DM flux signature in a conservative way. We assume no knowledge
about incidence energy and angle.  A much larger parameter region
can be probed if one applies the energy and angular information.

We compare the anti-DM flux induced event rate with the
indistinguishable background from the neutral-current interaction of
the atmospheric neutrino. We account for the neutrino fluxes of all
flavors above 100 MeV, since lower energy neutrino will not be able
to scatter with an individual nucleon but the whole nucleus of
oxygen. By requiring the elastic scattering rate from anti-DM flux
to be smaller than 2-$\sigma$ uncertainty of the rate from
atmospheric neutrino flux~\cite{Honda:2006qj,Honda:2011nf}, one can
constrain the rate of IND process happening in the Sun.

\begin{figure}[t]
\begin{center}
\hspace*{-0.75cm}
\includegraphics[width=0.50\textwidth]{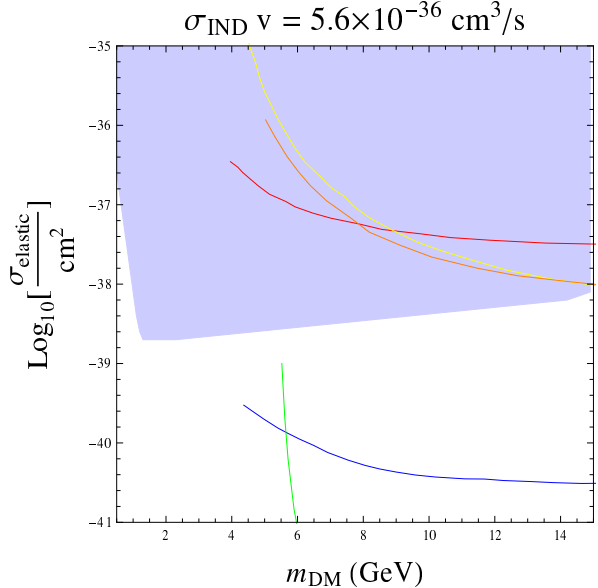} \hspace{0.1cm}
\caption{The constraint from anti-DM flux on the elastic scattering
cross section between cosmic DM and nucleus as a function of the DM
mass. Here we assume the IND interaction cross section to be the
value of our benchmark point. The solid lines indicate the
constraints from direct detection for spin-dependent and
spin-independent scattering. The red curve is from PICCASO, orange
is from SIMPLE, yellow is from COUPP, Blue is from CDMSLite and
Green is from
LUX.~\cite{Archambault:2012pm,Felizardo:2010mi,Behnke:2012ys,Akerib:2013tjd,Agnese:2013lua}}
\label{fig:AntiDMSun}
\end{center}
\end{figure}

In Fig.~\ref{fig:AntiDMSun}, we present constraints on the elastic
scattering cross section by anti-DM flux from the Sun, assuming the
IND interaction cross section as the value of the benchmark point.
Here we see that even without energy and angular information, the
anti-DM flux provides a reasonable probe of the parameter space. The
constraints on SD elastic scattering cross section is better than
the constraints from direct detections for a large mass range. It
also probes the very light DM mass region which is below the
threshold of the direct detections. Here we emphasize that if one
apply further energy and angular information from the fast moving
proton track, the atmospheric neutrino background can be efficiently
reduced, and we will be able to probe much larger parameter space.

Finally, one point needs to be addressed for the detection of the
anti-DM flux.  All the direct detection constraints on the elastic
scattering cross section is derived for the cosmic DM, whose
velocity relative to the nucleus is about $10^{-3}\ c$. However, the
anti-DM flux from the Sun has a much larger velocity comparing to
the cosmic DM.  In the previous study, we assume that the leading
order interaction cross section has no velocity dependence. If the
elastic scattering cross section between DM and nucleon has
non-trivial velocity dependence at leading order, e.g., $v^2$
\footnote{The different portals and their constraints are discussed
in detail in~\cite{MarchRussell:2012hi}. The collider bounds on
these operators, for example, $\mathcal{O}^{\phi}_{va} =
\frac{1}{\Lambda^2}\phi^{\dagger}\partial^{\mu}\phi\bar{f}\gamma_{\mu}\gamma^5
f$, may be evaded by introducing light mediators}, then the
fast-moving anti-DM from the Sun could have a much larger scattering
cross section than the direct detection bound. However, if the
elastic scattering cross section has a strong velocity dependence,
anti-DM from the IND process may not be able to leave the Sun
without colliding with the nucleons in the Sun. The elastic
scattering cross section between the fast moving anti-DM and nucleon
is required to be smaller than $8\times 10^{-37}\textrm{cm}^2$ in
order to escape the Sun.  If anti-DM cannot leave the Sun, then it
will be trapped and annihilate with the DM particle in the Sun. One
can instead constrain the model through the neutrinos in the final
state of dark matter pair annihilation. The detailed numbers are
model dependent, e.g. the neutrino branching ratio and its energy
spectrum, and we refer the reader to ~\cite{Hooper:2008cf} for a
detailed analysis.

\section{Conclusion}\label{SEC::Conclusion}

In this paper, we study a special scenario of asymmetric dark matter
model where DM particle carries anti-baryon and anti-lepton numbers.
Our model is inspired by hylogenesis
model,~\cite{Davoudiasl:2010am,Davoudiasl:2011fj,Blinov:2012hq}, but
we have several advantages. In original hylogenesis model, there are
two species of DM particles, one fermion and one boson. Their masses
need to be almost degenerate to avoid the decay between these two
species. In our model, we have a similar mechanism to generate
baryon asymmetry, but we have only one species of DM particle. Thus
our dark matter sector is simpler and no degeneracy is requested.
From the signature point of view, we have one lepton in our final
state which helps to improve the signature searches.

Since DM particle carries anti-baryon/lepton numbers, they can
annihilate with nucleons and induce striking signatures.  One of the
signatures is the induced proton decay signal in proton decay
experiments, such as SuperK. Similar signature also shows up in
hylogenesis model.  Because of the fact that we can have an
additional positron in the final state, our induced proton decay
process shares the same SM final states as the most sensitive search
channel in SuperK, i.e. $p^+\rightarrow e^+ + \pi^0$.  If we apply
the same event selection cuts as what are currently carried in
SuperK, one can probe interesting parameter region in very near
future.  We also give an example on how well one can probe our
parameter space by optimizing the event selection cuts respect to
our signature.  A mild change of cuts improve the sensitivity
dramatically, and current SuperK data has already been capable to
probe this model.

Further, if DM particles are captured by the Sun, the IND process
can induce large anti-neutrino and anti-DM fluxes. The neutrino
experiments can be used to study the IND process rate happening in
the Sun by constraining the anti-neutrino flux.  This can later be
interpreted as parameters in our model. Improving energy and angular
resolution of the neutrino experiments can largely enhance the
sensitivity.  As an illustration, we show how well such information
can help us studying our model by a reasonable choice of resolutions
according to ICANOE. Finally, anti-DM flux from the Sun can induce
similar signature as neutral current interaction between atmospheric
neutrinos and nucleons.  A conservative estimation is carried out to
show how well such signature can be used to probe our parameter
space.  A more detailed study taking into account the energy and
angular information of the fast moving proton in the detector can
further improve the sensitivity.

\begin{acknowledgments}
We would like to thank Josh Berger, Koun Choi, Savas Dimopoulos,
Peter Graham, Giorgio Gratta, Guey-Lin Lin, Jeremy Mardon, Michael
Peskin, Surjeet Rajendran, Michael Smy, Henry Sobel, Matt Strassler,
Yuhsin Tsai, Sean Tulin and Kathryn Zurek for helpful discussions.
This work was supported by ERC grant BSMOXFORD no. 228169 and NSF
grant PHY-1316699.
\end{acknowledgments}

\appendix
\section{DM pair annihilation and elastic scattering}
In the bulk of the paper, we focus mainly on an ADM model. However,
IND processes only rely on the fact that DM carries anti-baryon
numbers. In this appendix, we consider the symmetric DM scenario.
For IND process in SuperK experiment, the result only changes by a
factor of 2, since in the case of symmetric DM, only half of the DM
particles can annihilate with nucleons.  On the other hand, when
calculating the DM accumulation in the Sun, if the DM is symmetric,
one has to add the annihilation contribution to the evolution
equation.

One may expect that the DM/anti-DM annihilation always dominates
over induced nucleon decay process.  However, DM/anti-DM
annihilation rate is proportional to the product of DM and anti-DM
number densities while the IND interaction rate is proportional to
the product of DM number density and nucleon number density.  Thus
the IND process gains a large boost from the enormous nucleon
density in the center of the Sun.

\begin{figure}[htb]
\centering
\includegraphics[width=0.5\columnwidth]{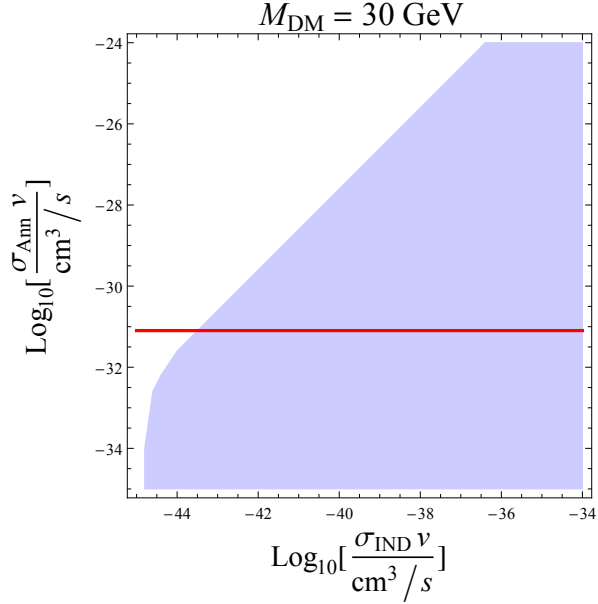}
\caption{For a symmetric DM scenario with DM mass as 30 GeV, this
figure shows how well the anti-neutrino flux from the Sun helps to
probe our parameter space.  The red line indicates the annihilation
cross section to give the correct relic
abundance.}\label{Fig:SymmetricDM}
\end{figure}

As an illustration to this point, we study a symmetric DM scenario
with DM mass of 30 GeV.  We assume the DM particle scatters with the
nucleon through spin-dependent interaction.  The elastic scattering
cross section is taken to be the largest value allowed by direct
detection experiment, i.e. $\sigma_{\mathrm{elastic}}=3\times
10^{-38}\textrm{cm}^2$.  Assuming the same energy and angular
resolution as stated in \ref{SEC::Neutrino} for ICANOE, we study how
well the anti-neutrino flux from the Sun can probe the
$\sigma_{\mathrm{annihilation}} - \sigma_{\mathrm{IND}}$ plane.  The result is shown
in Fig.~\ref{Fig:SymmetricDM}.

Here we see that a large region of parameter space can be probed by
the IND process induced anti-neutrino flux from the Sun, even for
symmetric DM scenario.  The red line on the plot is the annihilation
cross section which gives the correct relic abundance from a
standard thermal history.

\section{Dark matter Accumulation: General aspects}\label{SEC::GENERAL}
In this section, we review the general aspects of DM accumulation by
stellar objects.




In our model, the DM particles accumulated will be partly converted
into its anti-particle by interacting with nucleons in the stellar
objects. The total number of dark matter particles $\phi$ and
anti-particles $\overline{\phi}$ can be calculated using:

\begin{eqnarray}
\frac{\mathrm{d}N_{\phi}}{dt} &=& C_{\phi}
- A_{\phi}N_{\phi}N_{\overline{\phi}}
- B_{\phi}N_{\phi}
- E_{\phi}N_{\phi}
\nonumber \\
\frac{\mathrm{d}N_{\overline{\phi}}}{dt} &=& \phantom{C_\phi}
- A_{\phi}N_{\phi}N_{\overline{\phi}}
+ \epsilon_{\overline{\phi}}B_{\phi}N_{\phi}
- E_{\phi}N_{\overline{\phi}}
\label{Eq:neqyb}
\end{eqnarray}

Here, the $C_{\phi}$ is the DM capture rates, the $A_{\phi}$
describes the DM anti-DM annihilation, the $B_{\phi}$ describes DM
conversion into anti-DM, while $\epsilon_{\overline{\phi}}$ is the
chance that the converted $\overline{\phi}$ is captured by the stellar
object. The $A_{\phi}$ and $B_{\phi}$ can be well approximated by

\begin{eqnarray}
A_{\phi} &\simeq& (\sigma v)_{\mathrm{annihilation}} \big/
\left(4\pi r_{\phi,th}^3/3\right),\\
B_i &\simeq& (\sigma v)_{\mathrm{IND}} \;(\rho_c/m_n),
\end{eqnarray}
$m_n$ is the mass of the nucleon, and $r_{\phi, th}$ is the thermal
radius of the dark matter particles in the stellar objects.

\begin{equation}
r_{\phi,th} = \left( \frac{9T_c}{4\pi G\rho_cm_{\phi}}\right)^{1/2},
\end{equation}
$\rho_c$ is the mean baryon density in the center and $T_c$ is
center's temperature. In the case of the Sun, the thermal radius for
GeV mass DM can be expressed as:

\begin{equation}
r_{\phi,th} \simeq (5\times 10^9\,\mathrm{cm}) \left( \frac{3 \mathrm{GeV}}{m_{\mathrm{DM}}}\right)^{1/2}
\left(\frac{T_c}{1.5\times 10^7\mathrm{K}}\right)^{1/2}\left(\frac{1.5\times 10^5\mathrm{kg/m}^3}{\rho_c}\right)^{1/2}
\end{equation}


The capture rate of DM through elastic scattering with nuclei in the
Sun can be written as

\begin{eqnarray}
C_{\phi} &\simeq & 1.3 \times10^{25}\mathrm{s}^{-1}\left(\frac{\rho_{\mathrm{DM}}}{0.3\mathrm{
GeV/cm^3}}\right)\left(\frac{270\ \mathrm{km/s}}{\bar{v}}\right)\left(\frac{1\mathrm{GeV}}{m_{\mathrm{DM}}}\right)\nonumber \\
&\times & \left[\left(\frac{\sigma_{\mathrm{H}}}{10^{-40}\mathrm{cm^2}}\right)S(m_{\mathrm{DM}}/m_{\mathrm{H}})+1.1 \left(\frac{\sigma_{\mathrm{He}}}{16 \times 10^{-40}\mathrm{cm^2}}\right)S(m_{\mathrm{DM}}/m_{\mathrm{He}})\right].
 \label{Eq:CaptureRate}
\end{eqnarray}
with $\bar{v}$ being the local dark matter velocity,
$\sigma_{\mathrm{H}}$ and $\sigma_{\mathrm{He}}$ are the scattering
cross sections between Hydrogen/DM and Helium/DM, respectively. The
kinematic suppression function $S(x)$ is defined as:

\begin{equation}
S(x)=\left[\frac{A(x)^{3/2}}{1+A(x)^{3/2}}\right]^{2/3}
\end{equation}
where
\begin{equation}
A(x)= \frac{3 x}{2(x-1)^2}\left(\frac{v_{\mathrm{esc}}}{\bar{v}}\right)
\end{equation}
$v_{\mathrm{esc}} \simeq 617 \mathrm{km/s}$ is the escape velocity
of the Sun.


\end{document}